\documentclass[12pt]{article}
\usepackage{graphics,times}
\usepackage{epsf}
\usepackage{amsmath,amssymb,latexsym,float,algorithm,algorithmic}
\usepackage{enumerate}
\usepackage{listings}
\newtheorem{thm}{Theorem}[section]
\newtheorem{cor}[thm]{Corollary}

\newtheorem{prop}[thm]{Proposition}
\newtheorem{rem}[thm]{Remark}
\numberwithin{equation}{section}
\newcommand{\norm}[1]{\left\Vert#1\right\Vert}
\newcommand{\abs}[1]{\left\vert#1\right\vert}

\newcommand{\Complex}{\mathbb C}

\newcommand{\R}{\text{\fontshape{n}\selectfont I\kern-.42exR}}

\newcommand{\1}{\text{\fontshape{n}\selectfont 1\kern-.56exl}}

\newcommand {\qed}{\hspace*{\fill}\text{$\Box$}}

\begin{document}
\title{
{\bf A fast minimal residual solver for overlap fermions}
}

\author{Artan Bori\c{c}i\\
        {\normalsize\it Physics Department, University of Tirana}\\
        {\normalsize\it Blvd. King Zog I, Tirana-Albania}\\
        {\normalsize\it borici@fshn.edu.al}\\
\vspace{.2cm}
{\normalsize and}\\
\vspace{.2cm}
        Alban Allko\c{c}i\\
{\normalsize\it Computer Science Section, Polytechnic University of Tirana}\\
{\normalsize\it Mother Theresa Square, Tirana-Albania}\\
{\normalsize\it alban@fie.upt.al}\\
}

\date{}
\maketitle

\begin{abstract}
Computing quark propagators with overlap fermions requires
the solution of a shifted unitary linear system.
Jagels and Reichel have shown that for such systems
it is possible to construct a minimal residual algorithm
by short recurrences. The J\"ulich-Wuppertal group have found
this algorithm to be the fastest among overlap solvers.
In this paper we present a three-term recurrence for the
Arnoldi unitary process. Using the new recurrence we construct
a minimal residual solver which is the fastest
among all Krylov subspace algorithms considered so far for the
overlap inversion.
\end{abstract}

\pagebreak

\section{Introduction}

Lattice theories with chiral quarks provide an accurate tool for studying
the physics of strong interactions. The physical information of these theories
is contained in quark propagators, which are then combined to construct
meson, nucleon and even more complicated elementary particle propagators.
Quark propagator computations amount to solving large linear systems of the type:
\begin{equation}\label{lin_sys}
Dx = b
\end{equation}
where $D \in \Complex^{N\times N}$ is a dense matrix operator representing the
chiral Dirac operator on a regular four dimensional space-time lattice,
$x,b \in \Complex^N$ are the quark propagator and its source.

More than a decade ago two formulations of chiral fermion theories were discovered:
the domain wall fermions \cite{Ka92,FuSha95} and overlap fermions \cite{NaNe93}.
These apparently different formulations are closely related to each
other \cite{Borici_qcdna3_intro}. In particular, the truncated overlap variant of
domain wall fermions \cite{Borici_TOV} can be shown to be equivalent to overlap
fermions in four dimensions at any lattice spacing \cite{Borici_00}.
Therefore, for computational purposes it is sufficient to consider the Neuberger
Dirac operator, which is a shifted unitary matrix of the form \cite{Ne98}:
\begin{equation}\label{D_def}
D = c_1 \1 + c_2 V
\end{equation}
where
\begin{eqnarray}
V & = & D_W(D_W^*D_W)^{-1/2} \\
c_1 & = & (1+m)/2 \\
c_2 & = & (1-m)/2
\end{eqnarray}
Here $D_W$ is the Wilson-Dirac operator, which is a non-Hermitian sparse matrix.
It is easy to see why $V$ is a unitary matrix: starting from
the singular value decomposition of  $D_W=X\Sigma Y^*$, one gets $V=XY^*$.

Another way to look into the Neuberger operator is to use operators
$\Gamma_5=$diag$(I_{N/2},-I_{N/2})$
\footnote{We assume a Dirac spinor ordering compatible with the definition of
$\Gamma_5$.}
and $\hat{\Gamma}_5 = $sign$(H_W)$,
where $H_W=\Gamma_5D_W$ is the Hermitian Wilson-Dirac operator.
In this case it is easy to see that the overlap operator takes the form:
\begin{equation}\label{gamma_5_form}
D = c_1 \1 + c_2 \Gamma_5 \hat{\Gamma}_5
\end{equation}
This representation can be used to show that $D^*D$ commutes with $\Gamma_5$.
As a result, $D^*D$ is block diagonal in the chiral basis of $\Gamma_5$.
This observation has important applications: if we would like to solve linear systems
with coefficient matrix $D^*D$ we can use Conjugate Gradients (CG) algorithm,
which we know is optimal. Indeed, CG is very well suited for the molecular dynamics
algorithm which generates the ensemble of gauge fields in lattice QCD
with overlap fermions.

However, another important task is the computation of quark propagators
which we consider here.
This requires the solution of the linear systems of the type \ref{lin_sys}.
One can still use CG on normal equations (CGNE) but this method
may not be optimal in this case. In this paper we seek optimal solutions
of the above system in the Krylov subspace:
\begin{equation}\label{Krylov_subspace}
x \in {\mathcal K}_k = span\{b,Db,\ldots,D^{k-1}b\}
\end{equation}
It is well known that the generalized minimal residual method (GMRES) is
the optimal method that can be constructed in this subspace.
However, it is well known as well that the underlying
Arnoldi process produces long recurrences, a feature which limits the
benefits of the GMRES algorithm for large problems.
Nonetheless, Jagels and Reichel \cite{JagelsReichel94} found that for shifted
unitary systems it is possible to construct an Arnoldi process with short
recurrences. The method, the shifted unitary minimal residual algorithm (SUMR)
was shown by J\"ulich-Wuppertal group \cite{Wuppertal_II} to be the fastest linear
solver for quark propagator computations.

Here we propose an approach, which is based on a three-term recurrence
of the unitary Arnoldi process. This process is used to construct two optimal
Krylov subspace methods: the shifted unitary orthogonal method (SUOM) and
its geometric optimal counterpart SHUMR. We show that this algorithm converges
faster then SUMR for all lattices and quark masses considered in this paper.

The paper is organised as follows: section 2 describes the new recurrence
of the unitary Arnoldi process. Section 3 deals with the construction of
the SUOM and SHUMR algorithms. In section 4 we
compare directly these algorithms to SUMR and other Krylov subspace methods.
Conclusions are drawn in section 5.

\section{A three-term recurrence for the Arnoldi unitary process}

Our aim in this section is to construct an orthogonal basis of the
Krylov subspace \ref{Krylov_subspace} with as few operations as possible
which will then be useful for the solution of the linear system
\ref{lin_sys}. As a starting point we use the well-known
Arnoldi process \cite{Arnoldi}, which is essentially a modified
Gram-Schmidt process, given in Algorithm \ref{Arnoldi_alg}.
\begin{algorithm}[htp]
\caption{\hspace{.2cm} Arnoldi process}
\label{Arnoldi_alg}
\begin{algorithmic}
\STATE $q_1 = b / ||b||^2$
\FOR{$~k = 1, \ldots$}
    \STATE $w = V q_k$
    \FOR{$~j = 1, \ldots, k$}
       \STATE $h_{kj} = q_j^* w$
       \STATE $w := w - q_j h_{jk}$
    \ENDFOR
    \STATE $h_{k+1,k} = ||w||_2$
    \IF{$h_{k+1,k} = 0$} \STATE stop \ENDIF
    \STATE $q_{k+1} = w / h_{k+1,k}$
\ENDFOR
\end{algorithmic}
\end{algorithm}
After $k$ steps of this algorithm, the next unnormalized Arnoldi vector is given by:
\begin{equation}
{\tilde q}_{k+1} = V q_k - \sum_{j=1}^k q_j h_{jk}
\end{equation}
The right hand side contains a linear combination over all computed
Arnoldi vectors. This makes the overall complexity of the algorithm to grow
quadratically with $k$. Moreover, the computer memory grows linearly
with $k$, a requirement which is often prohibitive and thus undesirable
for large problems like ours.

Fortunately, for unitary matrices one can do better. After $k$ steps of
Algorithm \ref{Arnoldi_alg} we can write:
\begin{equation}\label{VQ_QH}
V Q_k = Q_k H_k + h_{k+1,k}  q_{k+1} e_k^T \equiv Q_{k+1} {\tilde H}_k
\end{equation}
where $Q_k = [q_1,\ldots,q_k]$ is the matrix of orthonormal
Arnoldi vectors, $H_k$ is an $k\times k$ upper Hessenberg matrix and
${\tilde H}_k$ is a $k+1\times k$ matrix obtained by appending the
row $h_{k+1,k} e_k^T$ to the matrix $H_k$.

\begin{prop}\label{H_orth}
${\tilde H}_k$ has orthonormal columns.
\end{prop}
\textbf{Proof}. Multiplying both sides of \ref{VQ_QH} from left by $Q_k^*$ we get:
\begin{equation}\label{H_k}
Q_k^* V Q_k = H_k
\end{equation}
Multiplying both sides of \ref{VQ_QH} from left again but now by $(VQ_k)^*$ we get:
\begin{equation}
I_k = Q_k^* V^* Q_k H_k + h_{k+1,k} (q_{k+1}^*VQ_k)^* e_k^T
\end{equation}
From \ref{H_k} and \ref{VQ_QH} one gets $Q_k^* V^* Q_k = H_k^*$ and
$q_{k+1}^*VQ_k = h_{k+1} e_k^T$. Thus, we find that:
\begin{equation}\label{orthonormal}
I_k = H_k^* H_k + h_{k+1,k}^2 e_k e_k^T = {\tilde H}_k^* {\tilde H}_k
\end{equation}
which proves the proposition. \qed

\begin{cor}\label{H_unitary}
From \ref{orthonormal} it is clear that $H_k$ is a
unitary matrix if its last column is normalised.
\end{cor}

This property was used be Jagels and Reichel to write $H_k$ as a product
of $k$ elementary Givens rotations, which are then exploited to construct
a coupled two-term recurrence Arnoldi algorithm (see Algorithm 3.1 of
\cite{JagelsReichel94}). They note that solving the coupled recurrences
would still lead to an algorithm where all vectors are required.

In contrast to their work, we give an algorithm which is defined by
short recurrences. We start by noting that the LU-decomposition of the
upper Hessenberg matrix $H_k$ can be written in the form:
\begin{equation}\label{LUm1}
H_k = L_k U_k^{-1}
\end{equation}
where $L_k$ is a lower bidiagonal matrix and $U_k$ is upper triangular.
For our convenience, we take the diagonal elements of $U_k$ to be one.
Substituting this decomposition into \ref{orthonormal}, multiplying the result
from the right by $U_k$ and from the left by $U_k^*$ we get:
\begin{equation}
U_k^*U_k = L_k^*L_k + h_{k+1,k}^2 e_k e_k^T
\end{equation}
Since the right hand side is a tridiagonal Hermitian matrix so must be
the left hand side. Hence, $U_k$ should be upper bidiagonal. This decomposition,
eq. \ref{LUm1} was used for the first time by Rutishauser to compute
the eigenvalues of orthogonal Hessenberg matrices \cite{Rutishauser66}.
We can easily extend this decomposition for the matrix $\tilde{H}_k$:
\begin{equation}
\tilde{H}_k = \hat{L}_k U_k^{-1}
, ~~~~\hat{L}_k = L_k + l_{k+1,k} e_k^T
\end{equation}
which gives $h_{k+1,k} = l_{k+1,k}$.

Multiplying both sides of \ref{VQ_QH} by $U_k$ we get:
\begin{equation}\label{VQU_QL}
V Q_k U_k = Q_k L_k + l_{k+1,k}  q_{k+1} e_k^T.
\end{equation}
This way, the next Arnoldi vector can be computed using:
\begin{equation}\label{three_term}
l_{k+1,k} ~q_{k+1} = (V - l_{kk} I) q_k + V q_{k-1} u_{k-1,k}
\end{equation}
where
\begin{equation}
u_{k-1,k} = - \frac{q_{k-1}^* V q_k}{q_{k-1}^* V q_{k-1}}
\end{equation}
and
\begin{equation}
l_{kk} = q_k^* V q_k + q_k^* V q_{k-1} u_{k-1,k}
\end{equation}
These expressions allow us to construct the three-term recurrence
Arnoldi unitary process as given in Algorithm \ref{AUP_alg}.
\begin{algorithm}[htp]
\caption{\hspace{.2cm} Arnoldi unitary process}
\label{AUP_alg}
\begin{algorithmic}
\STATE $u_{01} = 0$, $q_1 = b / ||b||^2$
\FOR{$~k = 1, \ldots n$}
    \STATE $w_k = V q_k$
    \STATE $u_{k-1,k} = - (q_{k-1}^* w_k) / (q_{k-1}^* w_{k-1})$
           $~~~~~~~~~$(for $k > 1$)
    \STATE $l_{kk} = q_k^* w_k + q_k^* w_{k-1} u_{k-1,k}$
    \STATE $w_{k+1} = w_k - l_{kk} q_k + u_{k-1,k} w_{k-1}$
    \STATE $l_{k+1,k} = ||w_{k+1}||_2$
    \IF{$l_{k+1,k} = 0$} \STATE stop \ENDIF
    \STATE $q_{k+1} = w_{k+1} / l_{k+1,k}$
\ENDFOR
\end{algorithmic}
\end{algorithm}

Note that the algorithm needs an additional inner product than the
usual (eg. Lanczos) three-term recurrence algorithms. The reason is the appearance
of the matrix $V$ both in the diagonal and subdiagonal terms of \ref{three_term}.
The algorithm is equivalent to the normal Arnoldi algorithm in exact arithmetic as
the following proposition shows:

\begin{prop}
Given $q_1$ and $V$ unitary the unitary Arnoldi process, Algorithm \ref{AUP_alg} 
generates the orthonormal vectors $Q_k$,
the lower and upper bidiagonal matrices $L_k$, $U_k$ such that
$\hat{L}_{k+1} U_k^{-1}$ is a matrix with orthonormal columns.
\end{prop}
\textbf{Proof}. The algorithm produces orthonormal vectors $Q_k$ and matrices
$L_k$ and $U_k$ such that equations \ref{VQU_QL} is satisfied. Multiplying both
sides of these equations by $U_k^{-1}$ we get new equations where $L_kU_k^{-1}$
is upper Hessenberg. From Proposition \ref{H_orth} follows that
$\tilde{H}_k$ and thus $\hat{L}_{k+1} U_k^{-1}$ is a matrix with orthonormal
columns. \qed

\begin{rem}
Both Arnoldi processes produce the same basis for the Krylov
subspace but Algorithm \ref{AUP_alg} is more efficient: its complexity is
linear in $k$ and its computer memory requirement is constant at each step $k$.
\end{rem}

\section{Optimal Krylov subspace methods for the overlap inversion}

The overlap operator is a non-Hermitian matrix operator. For such matrices
GMRES is knwon to be the optimal algorithm.
We call this class of algorithms geometrically optimal.
Another algorithm which can be used for such problems is the full orthogonalisation
method (FOM). In this case the $k^{\text{th}}$ residual vector,
\begin{equation}\label{res_def}
r_k = b - D x_k
\end{equation}
lies orthogonal to the Krylov subspace ${\cal K}_k$.
Because of this algebraic property we call this class of algorithms
algebraically optimal. It can be
shown that when the norm-minimising process of GMRES is converging rapidly,
the residual norms in the corresponding Galerkin process 
of FOM exhibit similar behaviour \cite{CullumGreenbaum96}.

Both methods GMRES and FOM use the Arnoldi process to generate the iterates.
In this section we will use the Arnoldi unitary process to construct two algorithms:
the first one is the specialisation of FOM to shifted unitary systems,
whereas the second is the specialisation of GMRES to these systems.

\subsection{A full orthogonalisation strategy}

We seek the solution in the Krylov subspace spanned by the Arnoldi vectors, i.e.
\begin{equation}
x_k = Q_k y_k, ~~~~y_k \in \Complex^k
\end{equation}
Assuming for simplicity that $\norm{b}=1$, and using eq. \ref{res_def}
the residual vector will be,
\begin{equation}
r_k = Q_k e_1 - D Q_k y_k
\end{equation}
Using the definition of D, eq. \ref{D_def} and the relation \ref{VQU_QL} we get:
\begin{equation}\label{res_Q}
r_k = Q_k e_1 - Q_{k+1} (c_1 \tilde{\1}_k + c_2 \hat{L}_k U_k^{-1}) y_k
\end{equation}
where $\tilde{\1}_k$ is obtained by appending a row of $k$ zeros to
the unit matrix $\1$. 
Projecting this equation onto the Krylov subspace ${\cal K}_k$ one has:
\begin{equation}
o = e_1 - (c_1 \1_k + c_2 L_k U_k^{-1}) y_k
\end{equation}
or equivalently:
\begin{equation}
(c_1 U_k + c_2 L_k) z_k = e_1, ~~~z_k = U_k^{-1} y_k
\end{equation}
Note that the matrix on the left hand side is tridiagonal and we can write:
\begin{equation}\label{T_k}
T_k z_k = e_1, ~~~T_k = c_1 U_k + c_2 L_k
\end{equation}
The LU decomposition of the matrix $T_k$ is denoted by
$T_k = {\tilde L}_k {\tilde U}_k$. Hence the solution can be written as:
\begin{equation}\label{z_k}
z_k = {\tilde U}_k^{-1}{\tilde L}_k^{-1} e_1 = {\tilde U}_k^{-1}
\begin{pmatrix}
{\tilde L}_{k-1}^{-1} e_1 \\
\alpha_k
\end{pmatrix}
=
\begin{pmatrix}
z_{k-1} \\
0
\end{pmatrix}
+ \alpha_k {\tilde U}_k^{-1} e_k
\end{equation}
where
\begin{equation}\label{alpha_k}
\alpha_k = e_k^T T_k^{-1} e_1 = e_k^T {\tilde L}_k^{-1} e_1
\end{equation}
Then from $x_k = Q_k U_k z_k$ with
$Q_k U_k = [Q_{k-1} U_{k-1}, q_k + q_{k-1} u_{k-1,k}]$ and from the
recurrence for $z_k$, eq. \ref{z_k} we get:
\begin{equation}
x_k = Q_{k-1} U_{k-1} z_{k-1} + \alpha_k ~Q_k U_k {\tilde U}_k^{-1} e_k
\end{equation}
Finally, denoting,
\begin{equation}\label{w_k}
w_k = Q_k U_k {\tilde U}_k^{-1} e_k
\end{equation}
we have:
\begin{equation}\label{x_k}
x_k = x_{k-1} + \alpha_k w_k
\end{equation}
Using this result and the definition of the residual vector, eq. \ref{res_def}
we get:
\begin{equation}\label{r_k}
r_k = r_{k-1} - \alpha_k D w_k
\end{equation}
Using matrices $\tilde{L}_k,\tilde{U}_k$ it is easy to show
(see Appendix A) that the following recurrences hold:
\begin{eqnarray}
\label{1st_rec}
w_k & = & q_k + q_{k-1} u_{k-1,k} +
\frac{\gamma_{k-1}}{{\tilde l}_{k-1,k-1}} w_{k-1} \\
\label{2nd_rec}
\alpha_k & = & \frac{\beta_{k-1}}{{\tilde l}_{kk}} \alpha_{k-1} \\
\label{3rd_rec}
{\tilde l}_{kk} & = &
 c_1 + c_2 l_{kk} - \frac{ \beta_{k-1} \gamma_{k-1} }{ {\tilde l}_{k-1,k-1} }
\end{eqnarray}
where $\beta_k = - c_2 l_{k+1,k}$ and $\gamma_k = - c_1 u_{k,k+1}$.

In this way we have specified the iterations for this linear solver,
which is called the shifted unitary orthogonal method or SUOM.
Below we give the MATLAB function SUOM.m.
\footnote{An e-copy of the function can be downloaded form the hep-lat
posting of the paper: MATLAB codes are left intact when included in
the body of the LaTeX source.}
The input is the right hand side vector $b$, the exact solution $x_0$,
the unitary matrix $V$, the real constants $c_1$ and $c_2$,
the tolerance $tol$ and the maximum number of iterations $imax$.
The output is the error norm history $rr$ and the solution $x$.

\vspace{0.5cm}
\begin{small}
\begin{lstlisting}
function [rr,x] = SUOM(b,x0,V,c1,c2,tol,imax);
b=b(:);
N=max(size(b));
vzero=zeros(N,1);
x=vzero;
r=b;

rho = norm(r);
rnorm=rho;
rr=norm(x0);
alpha = rho;
u12=0;
beta=1;
L11_tilde=1;
q=r/rho;
q_old=vzero; v_old=vzero; w_old=vzero; s_old=vzero;

counter = 1;
while ( (rnorm > tol) & (counter<=imax) );
  v=V*q;
  if (counter > 1),
    u12=-(q_old'*v)/(q_old'*v_old);
  end
  gamma=-c1*u12;
  L11=(q'*v)+u12*(q'*v_old);
  q_tilde=v-L11*q+u12*v_old;
  L21=norm(q_tilde);                                                      
  if (L21<=tol), break, end;                                      
  w=q+q_old*u12+w_old*gamma/L11_tilde;
  s=c1*(q+q_old*u12)+c2*(v+v_old*u12)+s_old*gamma/L11_tilde;
  L11_tilde=c1+c2*L11-beta*gamma/L11_tilde;
  alpha=alpha*beta/L11_tilde;  
                                                      
  x=x+w*alpha;
  r=r-s*alpha;

  q_old=q; v_old=v; w_old=w; s_old=s;                                         
  q=q_tilde/L21;
  beta=-c2*L21;
  rnorm=norm(r);
  rr=[rr;norm(x0-x)];
  counter++;
end
\end{lstlisting}
\end{small}

Note that the code can be trivially changed in order to get the
residual error norm instead of the error norm, or to use $x_0$ as a starting
guess by defining the starting residual error as $r = b - Dx_0$.

Another advantage of our MATLAB code is its easy inclusion into
a C++ code which uses {\tt uBLAS} libraries for inner products and Euclidean norms
\footnote{{\tt http://www.boost.org/libs/numeric/ublas/doc/overview.htm}}.
We have used these libraries to construct such a C++ function for use with
overlap fermions. In this case, the matrix-vector multiplication is an external routine which applies the inverse square root of $D_W^*D_W$ to the current
Arnoldi vector $q$ followed by a $D_W$ multiplication.

\subsection{A minimal residual strategy}

As in the previous case our starting point is the Krylov subspace ${\cal K}_k$.
But instead of projecting the residual vector we seek the
minimum of its 2-norm on this subspace. The solution in this case is denoted
by $\tilde{x}_k$ and is written formally as a solution of a
Least Squares Problem (LSP):
\begin{equation}\label{x_iterates}
\tilde{x}_k = \text{arg} \min_{x \in {\cal K}_k} \norm{b - D x}_2
\end{equation}
Requiring $x = Q_k y$ and using eq. \ref{res_Q} we get:
\begin{equation}\label{y_iterates}
\tilde{y}_k = \text{arg} \min_{y \in \Complex^k} 
\norm{Q_{k+1} ~[e_1 - (c_1 \tilde{\1}_k + c_2 \hat{L}_k U_k^{-1})y]~ }_2
\end{equation}
Since Arnoldi vectors are orthonormal, the matrix $Q_{k+1}$ can be ignored
and we end up with a much smaller LSP.
Note that the matrix ${\tilde H}_k = \hat{L}_k U_k^{-1}$ has orthonormal
columns (see Proposition \ref{H_orth}), a property which can be used to get
a short recurrence algorithm as in the case of the SUMR algorithm
\cite{JagelsReichel94}. However, as in the case of SUOM,
we follow a strategy that involves a tridiagonal matrix in the LSP.
This way the solution of the smaller problem is given by:
\begin{equation}\label{z_iterates}
\tilde{z}_k = \text{arg} \min_{z \in \Complex^k} 
\norm{e_1 - (c_1 U_k + c_2 \hat{L}_k)z}_2
, ~~~~\tilde{y}_k = U_k \tilde{z}_k
\end{equation}
where
\begin{equation}
c_1 U_k + c_2 \hat{L}_k =
\begin{pmatrix}
T_k \\
\nu e_k^T \\
\end{pmatrix}
\equiv \tilde{T}_k
, ~~~\nu = c_2 ~l_{k+1,k}
\end{equation}

In order to compute the iterates from those of the SUOM algorithm
we split the solution vector as follows:
\begin{equation}\label{splitting}
\tilde{z}_k = z_k + \xi_k
\end{equation}
Using equations \ref{T_k}-\ref{alpha_k} we get:
\begin{equation}
\xi_k = \text{arg} \min_{\xi \in \Complex^k} 
\norm{\nu \alpha_k e_{k+1} + \tilde{T}_k \xi}_2
\end{equation}

We solve this LSP by QR factorization of the matrix $\tilde{T}_k$:
\begin{equation}
\tilde{T}_k = O_k^* \tilde{R}_k
,~~~\tilde{R}_k =
\begin{pmatrix}
R_k \\
0 \\
\end{pmatrix} 
\end{equation}
with $O_k$ being a $k+1 \times k+1$ unitary matrix and $R_k$ an upper tridiagonal
matrix. Therefore we have:
\begin{equation}
\xi_k = \text{arg} \min_{\xi \in \Complex^k}
\norm{\nu \alpha_k O_k e_{k+1} + \tilde{R}_k \xi}_2
\end{equation}
As it is usual in this case,
the unitary matrix can be constructed using Givens rotations. At step $k$
one can express it in the form:
\begin{equation}\label{O_k}
O_k = G_k
\begin{pmatrix}
O_{k-1} & 0 \\
0 & 1 \\
\end{pmatrix}
,~~~~G_k =
\begin{pmatrix}
I_{k-1} & & \\
 & c_k & s_k \\
 & -\bar{s}_k & c_k \\
\end{pmatrix}
,~~~c_k^2 + \abs{s_k}^2 = 1
\end{equation}
From this it is clear that:
\begin{equation}\label{Oe_k+1}
O_k e_{k+1} = s_k e_k + c_k e_{k+1}
\end{equation}
which gives:
\begin{equation}
\min_{\xi \in \Complex^k} \norm{\nu \alpha_k O_k e_{k+1} + \tilde{R}_k \xi}_2
 = \min_{\xi \in \Complex^k} \norm{\nu \alpha_k s_k e_k + R_k \xi}_2
 + \abs{\nu \alpha_k c_k} 
\end{equation}
The right hand side is minimal if its first term is minimal.
If we assume that $R_k$ has full rank then this term must vanish.
In this case the solution is given by:
\begin{equation}
\xi_k = -R_k^{-1} e_k \nu \alpha_k s_k
\end{equation}
Using \ref{y_iterates}-\ref{splitting}, the solution to the original problem
can be written as:
\begin{equation}
\tilde{x}_k = Q_kU_kz_k + Q_kU_k\xi_k = x_k + Q_kU_k\xi_k \equiv x_k + \hat{x}_k
\end{equation}
where
\begin{equation}
\hat{x}_k = -Q_k U_k R_k^{-1} e_k \nu \alpha_k s_k
\end{equation}
In order to simplify the expression we define the matrix $P_k$:
\begin{equation}\label{P_k}
P_k = [p_1,\ldots,p_k] = Q_k U_k R_k^{-1}
\end{equation}
and denote $\omega_k = \nu \alpha_k s_k$. This way we have:
\begin{equation}\label{x_hat}
\hat{x}_k = -\omega_k p_k
\end{equation}
To complete the algorithm one should write the recurrence on $p_k$. Multiplying
both sides of \ref{P_k} by $R_k$ from the right we get $P_kR_k = Q_kU_k$, which
gives:
\begin{equation}
p_k \mu_k + p_{k-1} \varepsilon_k + p_{k-2}\theta_k = q_k + q_{k-1} u_{k-1,k}
\end{equation}
where by $\mu_k,\varepsilon_k$ and $\theta_k$ we denote the only non-zero entries
of the last column of $R_k$. From this equation we get:
\begin{equation}
p_k = (q_k + q_{k-1} u_{k-1,k} - p_{k-1} \varepsilon_k - p_{k-2}\theta_k) / \mu_k
\end{equation}
This completes the description of the method, which we call SHUMR in order to
distinguish it from the SUMR algorithm.
The details of the QR decomposition are given in Appendix B.
The MATLAB code of the algorithm is listed below.
It differs from the SUOM code with lines between
``{\small $start ~~added ~~lines$}'' and
``{\small $end ~~added ~~lines$}''. One can make here the
same remarks as made in the case of the SUOM.m function. In particular,
the code can be easily modified into a C++ code using {\tt uBLAS} libraries.

\vspace{0.5cm}
\begin{small}
\begin{lstlisting}
function [rr,x] = SHUMR(b,x0,V,c1,c2,tol,imax);
b=b(:);
N=max(size(b));
vzero=zeros(N,1);
x=vzero;
r=b;

rho = norm(r);
rnorm=rho;
rr=norm(x0);
alpha = rho;
u12=0;
beta=1;
L11_tilde=1;
q=r/rho;
q_old=vzero; v_old=vzero; w_old=vzero; s_old=vzero;

% start added lines
c_km1=1; s_km1=0; c_km2=0; s_km2=0;
p1=vzero; p2=vzero; Dp1=vzero; Dp2=vzero;
% end added lines

counter = 1;
while ( (rnorm > tol) & (counter<=imax) );
  v=V*q;
  if (counter > 1),
    u12=-(q_old'*v)/(q_old'*v_old);
  end
  gamma=-c1*u12;
  L11=(q'*v)+u12*(q'*v_old);
  q_tilde=v-L11*q+u12*v_old;
  L21=norm(q_tilde);
  if (L21<=tol), break, end;
  w=q+q_old*u12+w_old*gamma/L11_tilde;
  s=c1*(q+q_old*u12)+c2*(v+v_old*u12)+s_old*gamma/L11_tilde;
  L11_tilde=c1+c2*L11-beta*gamma/L11_tilde;
  alpha=alpha*beta/L11_tilde;
                                                                      
  x=x+w*alpha;
  r=r-s*alpha;
                                                                      
  q_old=q; v_old=v; w_old=w; s_old=s;
  q=q_tilde/L21;
  beta=-c2*L21;
  rnorm=norm(r);

% start added lines
  t11=c1+c2*L11;
  mu=t11*c_km1+gamma*conj(s_km1)*c_km2;
  nu=c2*L21;
  if (mu != 0),
    c_k=abs(mu)/sqrt(abs(mu)*abs(mu)+abs(nu)*abs(nu));
    s_k=conj(c_k*nu/mu);
  else
    c_k=0;
    s_k=1;
  end
  omega=nu*alpha*s_k;
  mu_k=c_k*mu+s_k*nu;
  eps=t11*s_km1-gamma*c_km1*c_km2;
  theta=-gamma*s_km2;
  p=(q+q_old*u12-p1*eps-p2*theta)/mu_k;
  Dp=(c1*(q+q_old*u12)+c2*(v+v_old*u12)-Dp1*eps-Dp2*theta)/mu_k;

  rnorm_p=norm(r+omega*Dp);
  xp=x-omega*p;

  c_km2=c_km1; s_km2=s_km1; p2=p1; Dp2=Dp1;
  c_km1=c_k; s_km1=s_k; p1=p; Dp1=Dp;
% end added lines

  rr=[rr;norm(x0-xp)];
  counter++;
end
\end{lstlisting}
\end{small}

\subsection{A numerical example}

We note that SUMR and SHUMR algorithms differ in the
underlying Arnoldi process: the SUMR algorithm uses two coupled
two-term recurrences which, when are solved, yield the usual long
recurrence of the Arnoldi algorithm; the SHUMR algorithm uses a three-term
recurrence. In principle, it is possible to compare theoretically the effect of
such a difference, but this goes beyond the purpose of this paper. We have chosen
instead a direct numerical comparison in case of overlap fermions,
since this is of great importance for practical lattice computations.
Before we do so we give a first example in the case of a small, i.e.
$200 \times 200$ unitary matrix.

The example is similar to Example 2 of the paper of Jagels and Reichel
\cite{JagelsReichel94}. Let $W$ unitary matrix of the order $200$
resulting from the QR decomposition of a matrix with random elements in the
interval $(0,1)$. Let $\Lambda$ be a diagonal unitary matrix with elements
$\lambda_k = e^{i\theta_k}$ where
\begin{equation*}
\theta_k = \pi (k-1)/6
, ~~~~1 \leq k \leq 6
\end{equation*}
and the rest is randomly distributed in the interval $(-\pi/4,\pi/4)$.
Then, the unitary matrix $V$ is defined by:
\begin{equation*}
V = W \Lambda W^*
\end{equation*}

\begin{figure}
\includegraphics{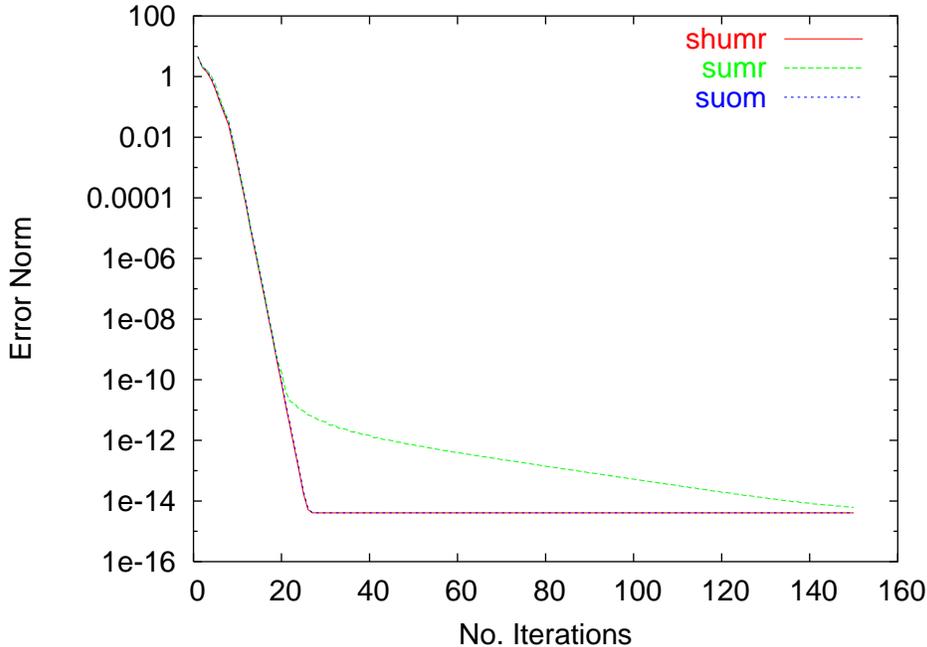}
\caption{Comparison of SHUMR,SUOM,SUMR algorithms for solving a small shifted
unitary linear system as described in the text.}
\label{small_example}
\end{figure}

We have tested SHUMR,SUOM,SUMR algorithms for solving the linear system
$Ax=b$ where $A=c_1\1+c_2V$ with $c_1=1.05$ and $c_2=1$ and random right hand
side $b$. The results are shown in Figure \ref{small_example}. We observe that
SHUMR and SUOM lie on top of each other (in this scale) and converge linearly
until they stagnate below $10^{-14}$. On the other hand the convergence rate of the
SUMR algorithm slows down when the value of the error norm is around $10^{-10}$.
It is not clear why SUMR differs in this way from SHUMR and SUOM in this
particular example.

\section{Comparison of algorithms for the overlap inversion}

In this section we compare the convergence of SUMR, SUOM and SHUMR algorithms
in the case of overlap fermions.
SUMR and SHUMR are geometrically optimal algorithms for shifted linear systems,
whereas SUOM is algebraically optimal in the sense defined above.
For completeness we display the convergence of Conjugate Residuals (CR),
Conjugate Gradients on Normal Equations (CGNE) and a special variation of
CGNE which we call CG-CHI. The latter exploits the fact that $D^*D$ is block
diagonal and solves simultaneously the two decoupled chiral systems.

Note that the computation of $D$ as applied to a vector is a numerical
problem, which is by now well researched.
A good review of these methods can be found in \cite{Wuppertal_I}.
We use the Lanczos method \cite{Borici_over,Borici_inv_sqr},
in the double pass version and without $H_W$-eigenvalue projection.

In the figures below we show the convergence of algorithms
as a function of Wilson matrix-vector multiplication number on
$8^316$ quenched lattices at various couplings and quark masses.

\begin{figure}
\includegraphics{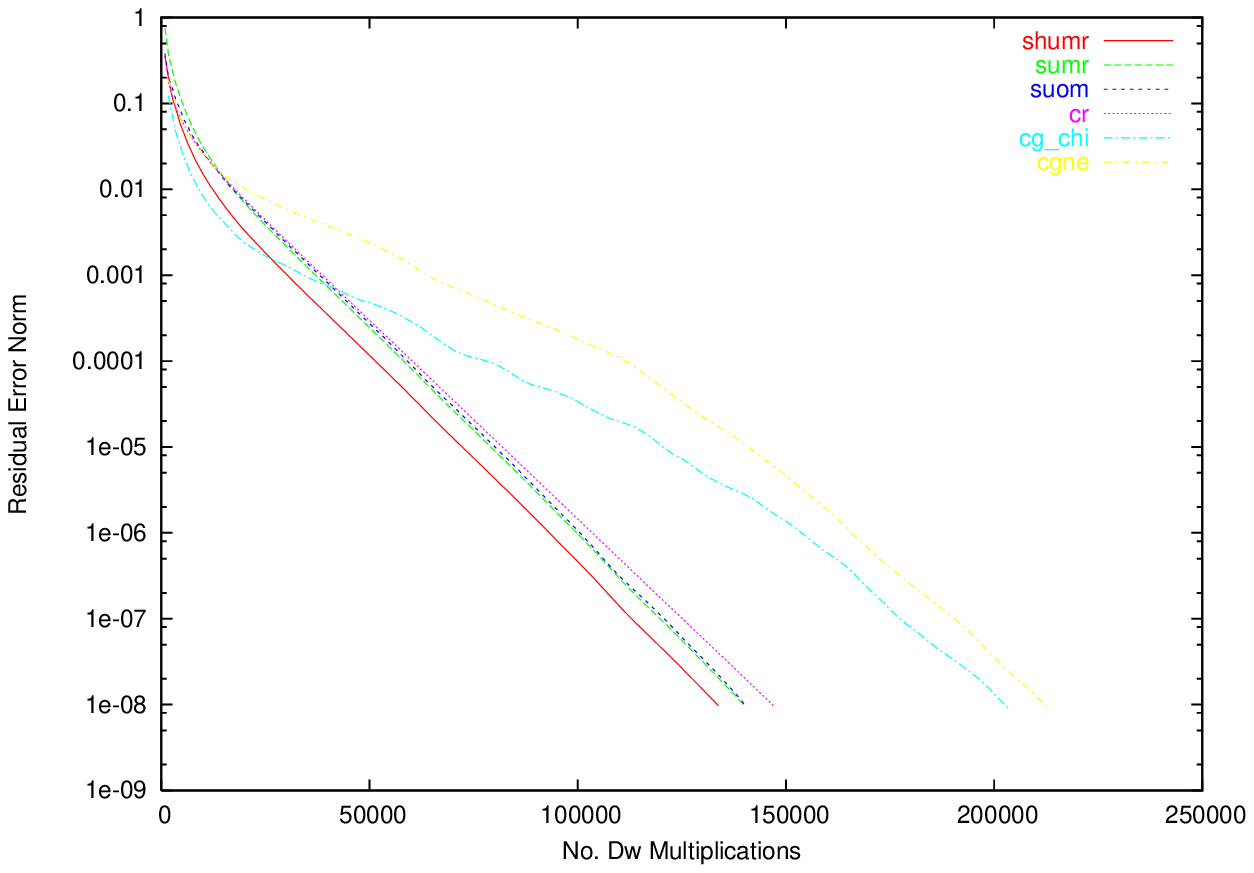}

\includegraphics{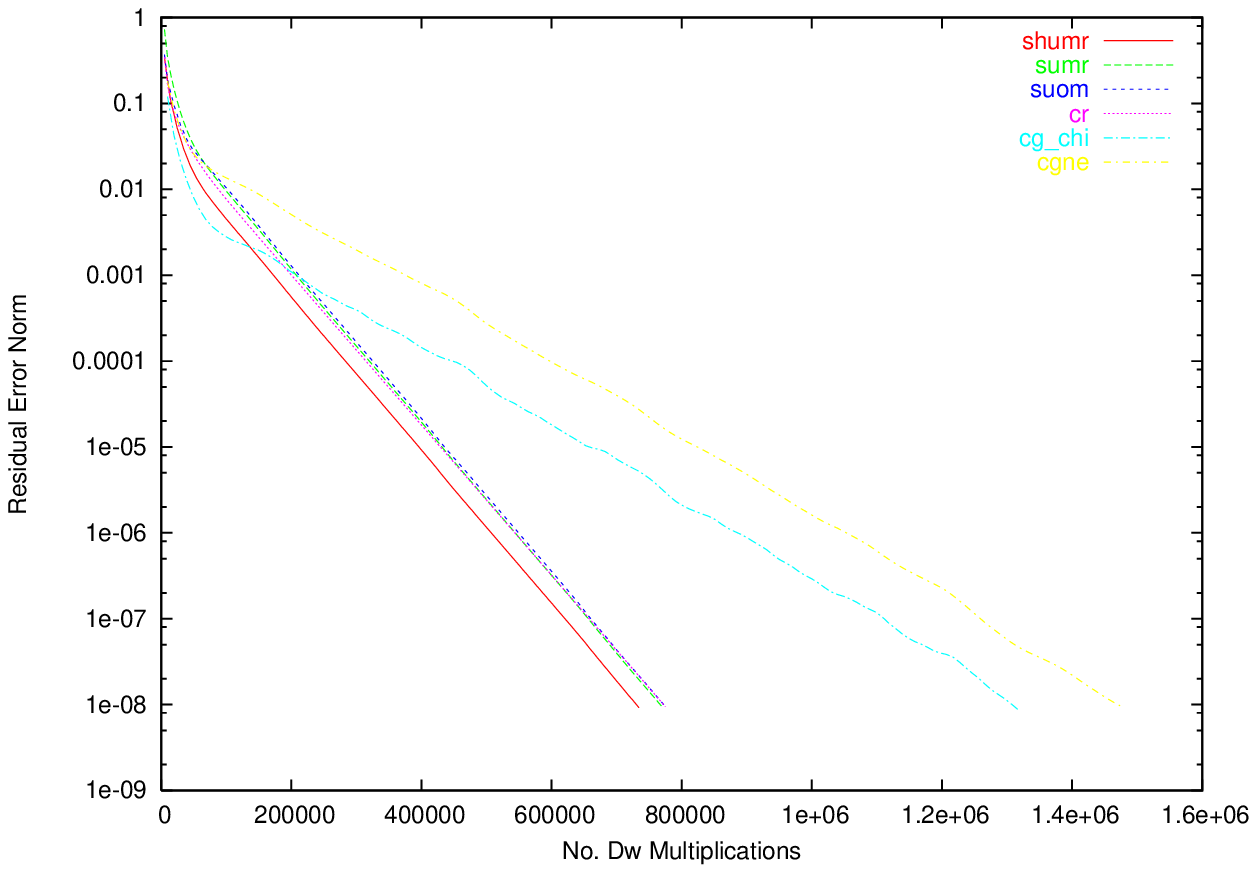}
\caption{Convergence history of various solvers on background gauge fields
at $\beta = 6$ (upper part) and $\beta = 5.7$ (lower part)
and quark mass $m=0.05$ }
\label{sssccc_mass_05}
\end{figure}

\begin{figure}
\includegraphics{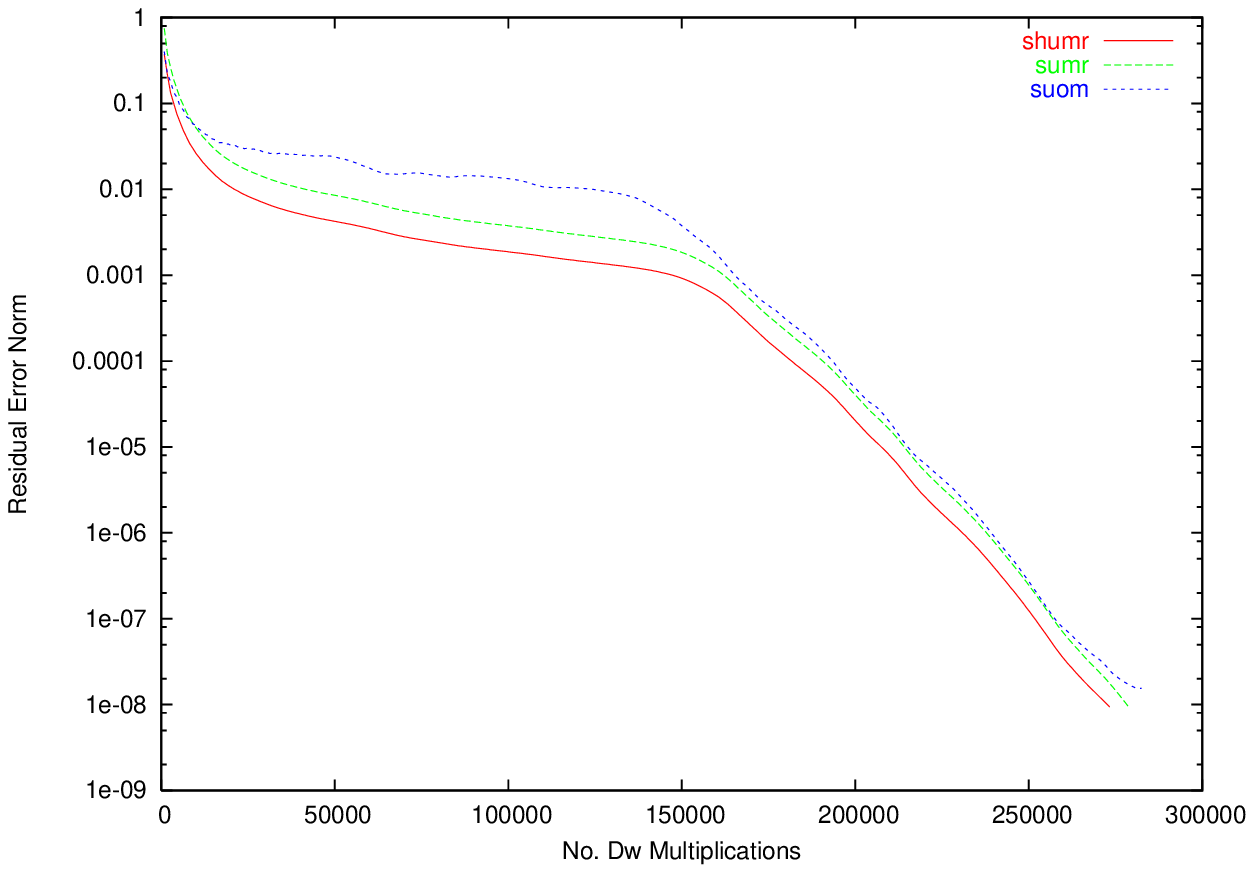}
\caption{Convergence history of SUMR, SUOM and SHUMR on background gauge fields
at $\beta = 6$ and quark mass $m=0.005$ }
\label{sss_mass_005}
\end{figure}

Figure \ref{sssccc_mass_05} compares all above mentioned algorithms
for quark mass $m=0.05$ at $\beta=6$ and $\beta=5.7$.
The first observation is that SHUMR, SUMR, SUOM and CR
are more efficient than CGNE and CG-CHI algorithms.
This is observed by the other groups as well \cite{Wuppertal_II}.
Hence, we decided not run these algorithms further for smaller quark masses.
The second observation is that SUMR, SUOM and CR converge neck-to-neck with
CR being slightly worse at $\beta=6$. The third observation is that SHUMR
converges $10-15\%$ faster than SUMR and SUOM.
\begin{figure}
\includegraphics{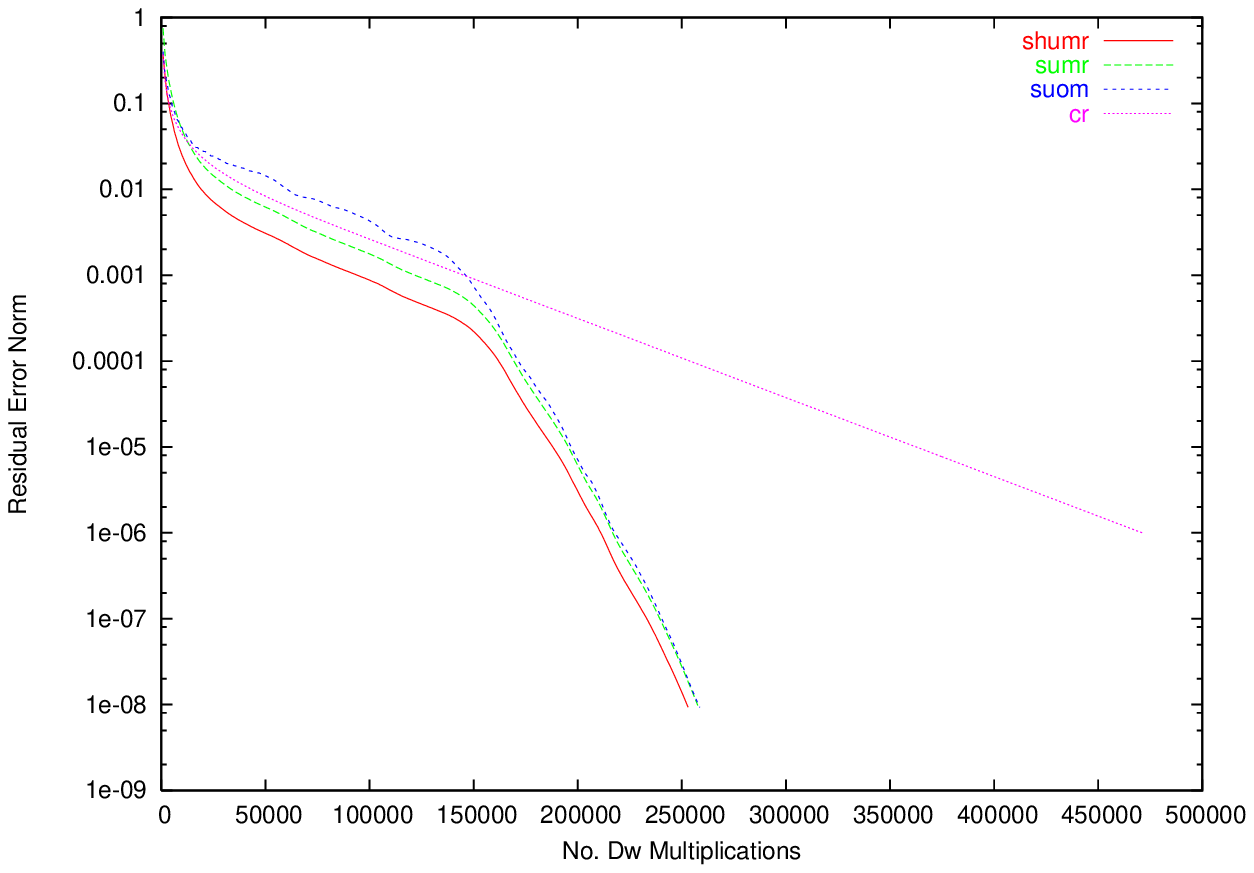}

\includegraphics{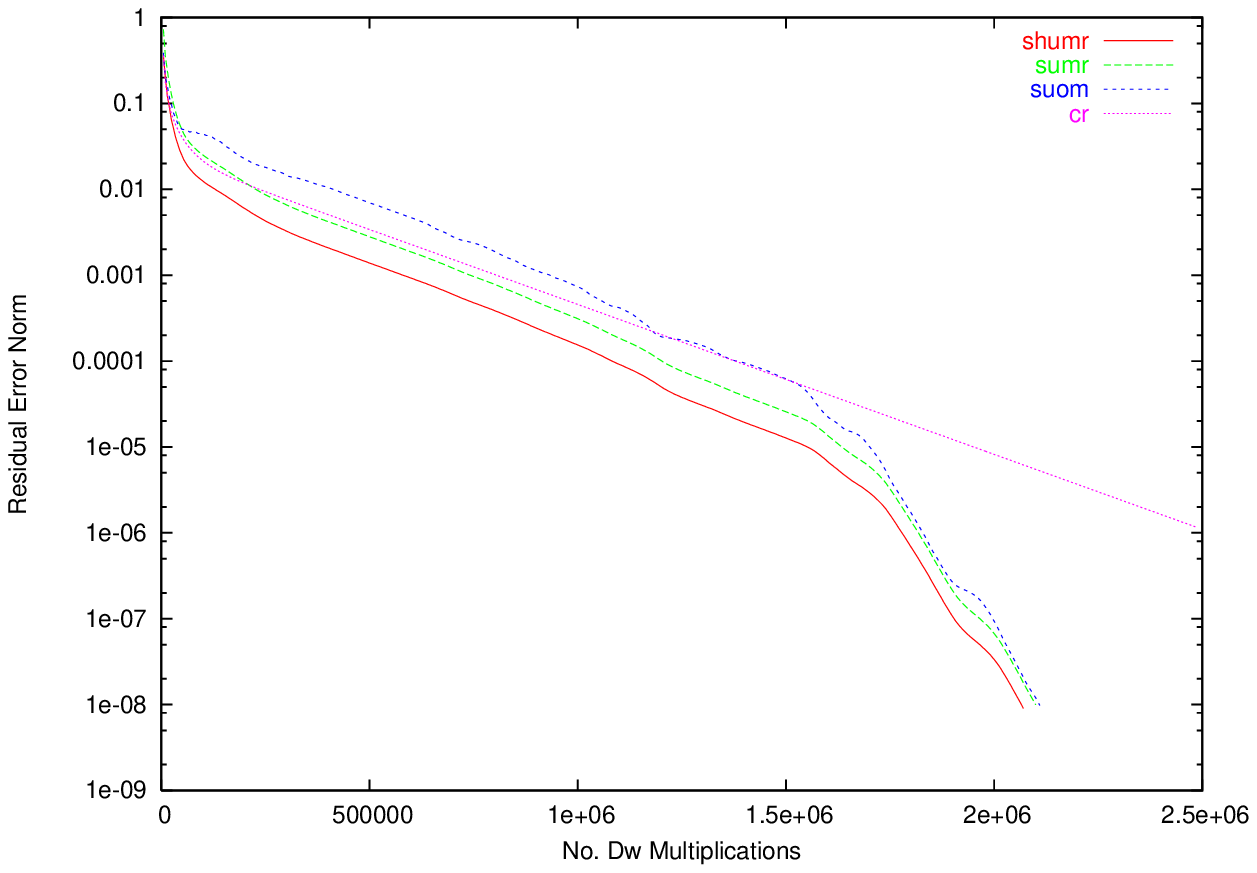}
\caption{Convergence history of SUMR, SUOM, SHUMR and CR on background gauge fields
at $\beta = 6$ (upper part) and $\beta = 5.7$ (lower part)
and quark mass $m=0.01$ }
\label{sssc_mass_01}
\end{figure}
Figure \ref{sssc_mass_01} compares the algorithms remaining in race
for the quark mass $m=0.01$ at $\beta=6$ and $\beta=5.7$. At this lighter mass
we observe a superlinear convergence rate of SHUMR, SUMR and SUOM: the rate
increases around residual norm $10^{-3}$ at $\beta=6$ and
around $10^{-5}$ at $\beta=5.7$. This is to be contrasted to the
linear convergence of CR. A second observation at this mass is the emergence
of the pattern that SHUMR converges faster than SUMR and the latter converges
faster than SUOM. 
This pattern is confirmed in Figure \ref{sss_mass_005} where the quark mass is
lowered to $m=0.005$.

Hence, the best algorithms are the optimal algorithms SHUMR and SUMR
with SHUMR converging $10-15\%$ faster in all cases.
Although both algorithms minimise the residual vector
norm they differ in the underlying Arnoldi process. The nature of short
recurrences employed by SHUMR may impact numerical properties of the algorithm.
Hence, generated Krylov subspaces are different and we may conclude that
SHUMR explores it more efficiently. However, at present we have no theoretical
tool to characterise the difference.

\section{Conclusions}

In this paper we have presented two iterative solvers which are based on a new
Arnoldi type algorithm for shifted unitary systems. The SUOM solver constructs
residual vectors which are orthogonal to the Krylov subspace. The SHUMR solver
minimises the residual vector over the generated Krylov subspace. Both algorithms
are short recurrence specialisations of FOM and GMRES for shifted unitary systems.
Taking into account the SUMR algorithm of Jagels and Reichel \cite{JagelsReichel94},
the short recurrence algorithms for such systems are hardly a new result.
But, it is somewhat
surprising that the SHUMR algorithm performs better than SUMR on the
examples shown in this paper. In particular, we presented a new Arnoldi unitary
process,
Algorithm \ref{AUP_alg}, which is intrinsically a short recurrence process,
a result which appears for the first time.

On the application level we conclude that SHUMR algorithm outperforms all other
knwon iterative solvers for quark propagator computations with overlap fermions.
The MATLAB codes of SUOM and SHUMR algorithms given here allow an easy conversion
to other object oriented code like C++ which uses {\tt uBLAS} libraries.

\pagebreak

\section*{Appendix A}

In order to derive
the recursions \ref{1st_rec}-\ref{3rd_rec} we partition the tridiagonal matrix
$T_k$ in the form:
\begin{equation*}
T_k =
\begin{pmatrix}
{\tilde L}_{k-1} {\tilde U}_{k-1}   &   -\gamma_{k-1} e_{k-1} \\
 - \beta_{k-1} e_{k-1}^T  & c_1 + c_2 l_{kk}
\end{pmatrix}
\end{equation*}
from where the LU factors are found to be:
\begin{equation*}
{\tilde L} =
\begin{pmatrix}
{\tilde L}_{k-1}                               &  0 \\
 - \beta_{k-1} e_{k-1}^T {\tilde U}_{k-1}^{-1} &  {\tilde l}_{kk}
\end{pmatrix},
~~~{\tilde U}_k =
\begin{pmatrix}
{\tilde U}_{k-1} & - \gamma_{k-1} {\tilde L}_{k-1}^{-1} e_{k-1} \\
 0               & 1
\end{pmatrix}
\end{equation*}
with
\begin{equation}\label{tilde_l_kk}
{\tilde l}_{kk} = c_1 + c_2 l_{kk}
 - \beta_{k-1} \gamma_{k-1} e_{k-1}^T T_{k-1}^{-1} e_{k-1}
\end{equation}
Their inversion gives:
\begin{equation*}
{\tilde L}_k^{-1} =
\begin{pmatrix}
{\tilde L}_{k-1}^{-1} &  0 \\
\frac{\beta_{k-1}}{{\tilde l}_{kk}} e_{k-1}^T {\tilde L}_{k-1}^{-1} &
\frac{1}{{\tilde l}_{kk}}
\end{pmatrix},
~~~{\tilde U}_k^{-1} =
\begin{pmatrix}
{\tilde U}_{k-1}^{-1} &
\frac{\gamma_{k-1}}{{\tilde l}_{k-1,k-1}} {\tilde U}_{k-1}^{-1} e_{k-1} \\
0 & 1
\end{pmatrix}
\end{equation*}

i) From \ref{w_k} and applying $e_k$ to the right of ${\tilde U}_k^{-1}$
one gets the first recursion of \ref{1st_rec}:
\begin{equation*}
w_k = q_k + q_{k-1} u_{k-1,k} +
\frac{\gamma_{k-1}}{{\tilde l}_{k-1,k-1}} w_{k-1}
\end{equation*}
where the initial vector is set to $w_1 = q_1$.

ii) Using \ref{alpha_k},
applying $e_k^T$ to the left and $e_1$ to the right of ${\tilde L}_k^{-1}$
one gets the second recursion of \ref{2nd_rec}:
\begin{equation*}
\alpha_k = \frac{\beta_{k-1}}{{\tilde l}_{kk}} \alpha_{k-1}
\end{equation*}
and with $\alpha_1 = \norm{b}_2 / \tilde{l}_{11}$.

iii) Finally, observing that:
\begin{equation*}
\frac{1}{ {\tilde l}_{kk} } = e_k^T T_k^{-1} e_k
\end{equation*}
and and using \ref{tilde_l_kk} gives the third recursion of \ref{3rd_rec}:
\begin{equation*}
{\tilde l}_{kk} =
 c_1 + c_2 l_{kk} - \frac{ \beta_{k-1} \gamma_{k-1} }{ {\tilde l}_{k-1,k-1} }
\end{equation*}
with $\tilde{l}_{11} = c_1 + c_2 l_{11}$.

\section*{Appendix B}

In order to complete the derivation of the SHUMR algorithm one has to specify
the Givens matrix $G_k$ and the last column of $R_k$. Let $\tilde{t}_k$ be the last
column of $\tilde{T}_k$ and $\tilde{\mu}_k$ be the last column of
$\tilde{R}_k$. Then, we have:
\begin{equation*}
\tilde{\mu}_k = O_k \tilde{t}_k
, ~~~~~~\tilde{t}_k = \nu e_{k+1} + t_{kk} e_k - \gamma_{k-1} e_{k-1}
\end{equation*}
where $t_{kk} = c_1 + c_2 l_{kk}$.
From the definition of $O_k$, eq. \ref{O_k} it is clear that:
\begin{equation}\label{Ot_k}
\tilde{\mu}_k = G_k
\begin{pmatrix}
t_{kk} O_{k-1} e_k - \gamma_{k-1} O_{k-1} e_{k-1} \\
\nu \\
\end{pmatrix}
\end{equation}
and from eq. \ref{Oe_k+1}:
\begin{equation}\label{O1}
O_{k-1} e_k = s_{k-1} e_{k-1} + c_{k-1} e_k
\end{equation}
Using \ref{O_k} again and applying the above result for $O_{k-2}e_{k-1}$ one finds:
\begin{equation*}
O_{k-1} e_{k-1} = G_{k-1}
\begin{pmatrix}
O_{k-2} e_{k-1} \\
0 \\
\end{pmatrix}
=
s_{k-2} G_{k-1} e_{k-2} + c_{k-2} G_{k-1} e_{k-1}
\end{equation*}
Since $G_{k-1} e_{k-2} = e_{k-2}$
and $G_{k-1} e_{k-1} = c_{k-1} e_{k-1} - \bar{s}_{k-1} e_k$ one gets:
\begin{equation}\label{O2}
O_{k-1} e_{k-1}
 = s_{k-2} e_{k-2} + c_{k-2} c_{k-1} e_{k-1} - c_{k-2} \bar{s}_{k-1} e_k
\end{equation}
Substituting \ref{O1} and \ref{O2} to \ref{Ot_k} one obtains:
\begin{equation*}
\tilde{\mu}_k = G_k
\begin{pmatrix}
\varepsilon_k e_{k-1} + \theta_k e_{k-2} \\
\mu \\
\nu \\
\end{pmatrix}
\end{equation*}
where
\begin{eqnarray*}
\mu & = & t_{kk} c_{k-1} + \gamma_{k-1} c_{k-2} c_{k-1} \\
\varepsilon_k & = & t_{kk} s_{k-1} - \gamma_{k-1} c_{k-2} c_{k-1} \\
\theta_k & = & - \gamma_{k-1} s_{k-2} \\
\end{eqnarray*}
with $c_0 =1$, $s_0 = c_{-1} = s_{-1} = \gamma_0 = 0$.

The values of $\mu_k$ and $s_k, c_k$ are determined by the condition:
\begin{equation*}
G_k
\begin{pmatrix}
\varepsilon_k e_{k-1} + \theta_k e_{k-2} \\
\mu \\
\nu \\
\end{pmatrix}
=
\begin{pmatrix}
\varepsilon_k e_{k-1} + \theta_k e_{k-2} \\
\mu_k \\
0 \\
\end{pmatrix}
\end{equation*}
or equivalently by:
\begin{equation*}
\begin{pmatrix}
      c_k   & s_k \\
 -\bar{s}_k & c_k \\
\end{pmatrix}
\begin{pmatrix}
\mu \\
\nu \\
\end{pmatrix}
=
\begin{pmatrix}
\mu_k \\
0 \\
\end{pmatrix}
\end{equation*}
It is easy to see that $\mu_k = c_k \mu + s_k \nu$ and $\bar{s}_k = c_k \nu / \mu$.
Using $c_k^2 + \abs{s_k}^2 = 1$ one has:
\begin{equation*}
c_k = \frac{\abs{\mu}}{\sqrt{\abs{\mu}^2 + \abs{\nu}^2}}
\end{equation*}
For $\mu = 0$ one has $c_k = 0$ and $s_k = 1$.

\end{document}